\documentstyle[psfig,11pt,aaspp4,flushrt]{article}  
\def\etal{{\it et al. }}
\def\kms{km~s$^{-1}$~}
\def\hal{H$\alpha$}

\def\r83{$r_{\rm 83}$}

\def\be{\begin{equation}}
\def\ee{\end{equation}}

\def\W50{W$_{50}$~}
\font\eightrm=cmr8
\font\eightbf=cmbx8
\font\eightit=cmti8
\font\eightsl=cmsl8
\font\eightmus=cmmi8
\def\smalltype{\let\rm=\eightrm \let\bf=\eightbf
  \let\it=\eightit \let\sl=\eightsl \let\mus=\eightmus
  \baselineskip=9.5pt minus .75pt
  \rm}

\def \nin+{584~}
\def \nin{374~}
\def \nin_use{346~}

\begin{document}
\title{The Inner Scale Length of Spiral Galaxy Rotation Curves}

\author {Riccardo Giovanelli, Martha P. Haynes}
\affil{Center for Radiophysics and Space Research
and National Astronomy and Ionosphere Center
\altaffilmark{1},\nl
Space Sciences Bldg., Cornell University, Ithaca, NY 14853\nl 
{\it e--mail:} riccardo@astro.cornell.edu,  haynes@astro.cornell.edu}

\altaffiltext{1}{The National Astronomy and Ionosphere Center is
operated by Cornell University under a cooperative agreement with the
National Science Foundation.}

\hsize 6.5 truein
\begin{abstract}
We use the tapering effect of \hal/[\ion{N}{2}] rotation curves of spiral
galaxies first noted by Goad \& Roberts (1981) to investigate the
internal extinction in disks. The scale length of exponential fits to the 
inner part of rotation curves depends strongly on the disk axial ratio.
Preliminary modelling of the effect implies substantial opacity of the 
central parts of disks at a wavelength of 0.66 $\mu$. In addition, the 
average kinematic scale length of rotation curves, when corrected to face--on 
perspective, has a nearly constant value of about $1.7h^{-1}$ kpc, for all 
luminosity classes. The interpretation of that effect, as the result of the 
increasing dominance of the baryonic mass in the inner parts of galaxies, 
yields a mean baryonic mass--to--light ratio in the I band
$\bar\Upsilon_I= 2.7h$ ~$M_\odot/L_{\odot,I}$, within the inner $1.7h^{-1}$ 
kpc of disks.

\end{abstract}


\section {Introduction}

At optical and near infrared wavelengths, interstellar dust depresses
the observed flux of galaxy disks by both scattering and absorption.
Several authors have proposed that, even in the face--on perspective, 
normal, non--starburst disks are optically thick at optical wavelengths, 
while others have argued for substantial transparency (see the volume
edited by Davies \& Burstein 1995 and the review by Calzetti 2001 for
details). Using a sample of spiral 
galaxies within $cz\sim 10,000$ \kms, we statistically derived photometric
solutions for the degree of internal extinction at I--band as a function
of disk inclination (Giovanelli \etal 1994), finding a difference of 
more than a magnitude of flux between face--on and edge--on systems. 
At the same time, we found evidence for transparency of the outer parts
of disks, at radii $>3$ disk scale lengths from the center. We later
reported that the amount of internal extinction is luminosity dependent:
more luminous disks being more opaque than less luminous ones 
(Giovanelli \etal ~1995). These results have been confirmed by more
recent analyses (e.g. Tully \etal ~1998; Wang \& Heckman 1996).

Purely photometric techniques are subject to a peculiar set of selection
effects, that can severely affect quantitative conclusions on internal
extinction, as witnessed by the liveliness of the debate over the last
decade. In 1981, Goad \& Roberts noted a kinematic effect which can
provide an independent test for disk extinction.

Let $V(r)$ be the rotation curve of a disk ($r$ being the radial
coordinate along the disk's projected major axis), assumed to have axial
symmetry. Let $(x,y)$ be a set of Cartesian coordinates {\it in the plane 
of the disk}. If the disk is thin and it is
observed at inclination $i$ ($i=90^\circ$ for edge--on), the component
of velocity along the line of sight which intercepts the disk at $(x,y)$
is  
\be
V_\parallel = V(r) {x\over \sqrt{x^2+y^2}} \sin i + V_{turb}, 
\ee 
where $V_{turb}$ accounts for turbulence and $x$ is oriented along the
disk's apparent major axis. An observed rotation curve, as derived for 
example from a long--slit \hal ~spectrum positioned along $x$, is 
smeared by seeing, instrumental resolution, averaging across the slit 
width, the finite thickness of the disk and extinction occurring within 
the disk itself. As realistically thick disks approach the edge--on perspective, 
lines--of--sight along the major axis sample regions of increasingly
broad range in $y$, yielding a velocity distribution with a peak velocity
contributed by parcels of gas at $y=0$ and a low velocity wing contributed
by parcels at $|y|>0$. If extinction is important, only foreground parts of
the disk contribute to the emission and the factor $(x/ \sqrt{x^2+y^2})<1$
depresses the velocity distribution observed at $r=x$. 
Goad \& Roberts noted how this tapering effect may, in opaque edge--on
disks, produce observed rotation curves resembling solid--body behavior,
independently of the true shape of $V(r)$.
Bosma \etal (1992) applied this technique to two edge--on systems, 
NGC 100 and NGC 891, by comparing HI synthesis and \hal ~observations.
The evidence led them to conclude that the disk of NGC 100 is transparent,
while in the case of NGC 891 they could not exclude the possibility of
extinction in the inner parts of the disk. Prada \etal (1994) compared
long--slit spectra of NGC 2146 in the optical and near IR and reported
evidence for extinction in the inner parts of the galaxy.

Here, we apply the Goad \& Roberts test in a statistically convincing 
manner to a sample of more than 2000 \hal/[\ion{N}{2}] rotation curves. Our results 
clearly indicate substantial opacity in the inner disks of spiral disks. 
They also show the effect to be luminosity dependent, in a manner very 
much in agreement with our previous photometric determination. In 
Section 2 we present our data sample and describe our
rotation curve fits. Results on extinction are discussed in Section 3.
In Section 4, we discuss an interesting serendipitous finding: the 
apparent constancy of the mean value of the kinematic scale length of 
\hal/[\ion{N}{2}] rotation curves, across populations of different
luminosity class. Throughout this paper, distances and luminosities are 
obtained from redshifts in the Cosmic Microwave Background 
reference frame and scaled according to a Hubble parameter 
$H_\circ = 100h$ \kms ~ Mpc$^{-1}$.

\section{Data Sample and Rotation Curve Model Fits}

As an effort to map the peculiar velocity field of the local Universe, 
we have assembled a sample of several thousand observations of I--band
photometry and rotational width data of spiral galaxies. Our observations, 
including those listed in Dale \& Giovanelli (2000 and refs. therein) as
well as several hundred obtained at the Hale 5 m telescope after that report
and to be presented elsewhere, have been complemented by the samples observed 
in the southern hemisphere by Mathewson and co--workers (Mathewson
\etal ~1992; Mathewson \& Ford 1996). The kinematic information for a 
substantial subset of these data is in the form of rotation curves 
in electronic form, derived from long--slit \hal/[\ion{N}{2}] spectra. 

Rotation curves can be fitted by a variety of parametric models, some
of which are motivated by the physical expectation of contributions
by a baryonic component with a mass distribution mimicking that of the
light plus a dark matter spheroidal halo. Other
models rely purely on the versatility of a mathematical form in fitting
effectively the observed rotation curves with a minimum of free parameters.
Here we report on fits to 2246 rotation curves with a parametric model of 
the form
\be
V_{pe}(r) = V_\circ (1-e^{-r/r_{pe}})(1+\alpha r/r_{pe}) 
\ee
where $V_\circ$ regulates the overall amplitude of the rotation
curve, $r_{pe}$ yields a scale length for the inner steep rise and $\alpha$
sets the slope of the slowly changing outer part. This simple model, which
we refer to as {\it Polyex}, has been found to be very ``plastic''. It fits 
effectively both the steeply rising inner parts of rotation curves, as well 
as varying outer slopes, and it can be used to advantage in estimating 
velocity widths at specific radial distances from the galactic center as
required by applications of the luminosity--linewidth relation. In particular, 
it provides a very useful estimate of the inner slope of the rotation curve 
as given by $r_{pe}$. Of the 2246 rotation curves fitted by the model shown 
in Equation (2), 425 have been rejected due to poor quality of the data or 
of the fit or because $r_{pe}$ projects to less than 3" and may thus be affected 
by poor angular resolution. Mean values of $\alpha$, the value of $r_{pe}$ as 
seen in face--on systems, $r^\circ_{pe}$, and of the ratio between $r_{pe}$ and 
the scale length of the disk light for face--on systems are given in Table 1, 
for different luminosity bins; mean values of $r_{pe}$ are shown in Figure 1 
and discussed in the next section, where they are used to probe disk extinction.
Contents of columns 5 and 6 of Table 1 are discussed in Section 4.

\begin{deluxetable}{cccccc}
\tablewidth{0pt}
\tablenum{1}
\tablecaption{Parameter Averages by Luminosity Class}
\tablehead{
\colhead{Luminosity Range}   & \colhead{$100<\alpha>$} &
$<h r^\circ_{pe}>     $          & $<r^\circ_{pe}/r^\circ_d>$ &
$<h r^\circ_{pe}>_1$\tablenotemark{a}  &  $<a^\circ_{pe}>_1$\tablenotemark{(a)}
\nl
 & & {\it kpc} & & {\it kpc} & {\it arcsec}
}
\startdata
$M-5\log h<-22.0$       & $-0.31\pm0.12$ & $1.86\pm0.09$ & $0.63\pm0.02$ & $1.90\pm0.11$ & $7.1\pm0.5$ \nl
$-22.0<M-5\log h<-21.4$ & $+0.23\pm0.11$ & $1.75\pm0.06$ & $0.68\pm0.02$ & $1.71\pm0.06$ & $6.2\pm0.3$ \nl
$-21.4<M-5\log h<-21.0$ & $+0.53\pm0.19$ & $1.73\pm0.09$ & $0.83\pm0.03$ & $1.84\pm0.11$ & $7.1\pm0.5$ \nl
$-21.0<M-5\log h<-20.4$ & $+1.64\pm0.17$ & $1.73\pm0.08$ & $0.95\pm0.03$ & $1.82\pm0.07$ & $7.8\pm0.5$ \nl
$-20.4<M-5\log h<-19.5$ & $+1.97\pm0.22$ & $1.63\pm0.09$ & $1.02\pm0.04$ & $1.85\pm0.09$ & $8.3\pm0.5$ \nl
$-19.5<M-5\log h$       & $+4.21\pm0.76$ & $1.50\pm0.15$ & $1.28\pm0.08$ & $1.94\pm0.14$ & $11.7\pm0.8$ \nl
\enddata
\tablenotetext{(a)}{Averages computed for $hr_{pe}> 1$ kpc, $a_{pe}> 3"$, distances 
between 20 and 80 $h^{-1}$ Mpc, and $\log (a/b)$ less than 0.3, 0.4, 0.5,
respectively for $M-5\log h<-21.0$, $<-19.5$ and $> -19.5$.}
\end{deluxetable}

\section{Evidence for Extinction}

In Figure 1, the mean value of $r_{pe}$ is shown separately for different
luminosity classes, as a function of the disk inclination as expressed by 
$\log (a/b)$, where $(a/b)$ is the apparent axial ratio. Each symbol is
the average within a bin including between 20 and 30 galaxies. The 
horizontal dotted line in each panel indicates an arbitrary scale of 
$1.6 h^{-1}$ kpc. The total 
number of galaxies and the absolute magnitude range is indicated within 
each panel of the Figure. The parameter $r_{pe}$ gives an indication of 
the radial distance at which the rotational velocity is $1-1/e=0.63$ of 
$V_\circ (1+\alpha)$ (the asymptotic velocity for flat rotation curves,
for which $\alpha=0$). Figure 1 shows that, as $\log (a/b)$ approaches 0.45
($i \simeq 70^\circ$), the kinematic scale length $r_{pe}$ of observed 
rotation curves starts increasing. The effect is more marked for more 
luminous disks, for which edge--on systems exhibit an average value of 
$r_{pe}$ nearly three times as large as that of systems with $i<45^\circ$. The 
effect is not evident in the least luminous systems (those with $M_I-5\log h$ 
fainter than -19.5). This result is in agreement with the expectations 
based on the photometric determination of Giovanelli \etal (1995). No
axial ratio dependence is observed for $\alpha$, the outer slope of the
rotation curves.

\begin{figure}[t]
\figurenum{1}
\centerline{\psfig{figure=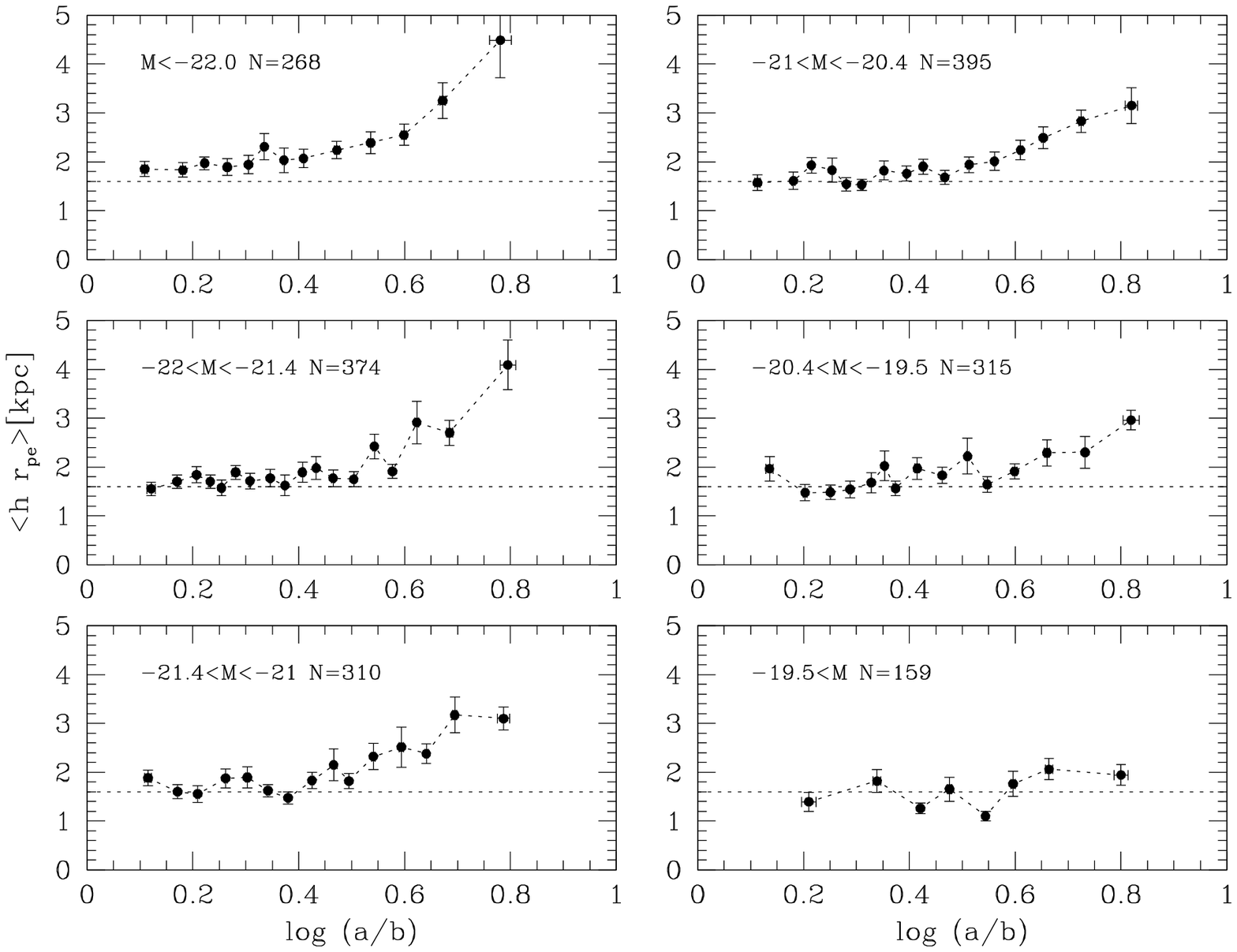}}
\caption{Average values of the kinematic scale length $h r_{pe}$,
plotted versus $\log (a/b)$, where $a/b$ is the galaxy's axial ratio.
Each panel displays data for the range in absolute magnitude ($M-5\log h$)
indicated in the figure, where $N$ is the total number of galaxies
included in each panel. Systems for which $r_{pe}$ projects at an angular
size $<3"$ have been excluded, in order to avoid angular resolution
bias. The dotted line in each panel is arbitrarily drawn 
at a constant value of 1.6 kpc.}
\end{figure}

We have carried out preliminary modelling of the observations shown in 
Figure 1 by simulating the dust and \hal ~distributions as exponential 
disks both in $r$ and $z$. Assuming the same distribution for dust and
\hal ~tracers, with a ratio between scale height and scale length of 
0.06 (as found by Xilouris \etal 1997; 1998), we obtain approximate 
match between models and Figure 1 for values of the the face--on 
($i=0^\circ$), central ($r=0$) optical depth of the model at the 
wavelegth of \hal ~($\lambda=0.66\mu$), $\tau_\alpha^\circ(0)$, 
of about 4. These modelling results are preliminary and depend very 
sensitively on the assumed thickness of disks and on the relative 
scales of dust and \hal ~sources. We will report more extensively
on the results of this effort at a later stage.

\section{A Characteristic Kinematic Scale Length}

Figure 1 shows that, for low axial ratios (where the effect of internal 
extinction is negligible), the average value of the kinematical scale 
length $r_{pe}$ is approximately the same for disks of all luminosities. 
Is this result to be expected?

First, we exclude that the angular resolution of spectroscopic observations 
has a significant impact on the reported findings. Not using galaxies for 
which $r_{pe}$ projects to an angular size $a_{pe}<3"$ rules out bias due 
to extreme cases of poor angular resolution. Since we use rotation curves from
different sources, which have been produced with different angular sampling
intervals, we have also verified that there is no systematic difference in
the fitted values of $a_{pe}$ among different data sources, taking advantage 
of significant sample overlaps. We have done so for several dozen galaxies 
appearing in at least two samples among those: of our own, of Mathewson et 
al., of Courteau (1997) and of Rubin et al. (1999 and refs. therein).

Second, we explore the effect of sample selection criteria. 
The values of $<r_{pe}>$ displayed in Figure 1 are affected by Malmquist
bias, which does not impact on the trends produced by extinction but alters
the relative values of $<r_{pe}>$ among different luminosity classes: 
since more luminous populations have larger average distance, for them the 
angular size limit of 3" produces an $<r_{pe}>$ biased high. In order to 
remove this bias, we draw a subsample which is volume limited, by including 
in it only systems with $a_{pe}>3"$, $hr_{pe}> 1$ kpc and within the distance 
range $20 h^{-1}$ to $80 h^{-1}$ Mpc. At $80 h^{-1}$ Mpc, an angular size 
limit of 3" translates into a linear size limit of about 1 kpc; the reason 
for a lower limit in distance is to reduce the uncertainty introduced
by peculiar velocities. We further restrict the subsample to low
inclination disks ($\log (a/b) < 0.3$, $<0.4$, $<0.5$ respectively for 
$M-5\log h<-21.0$, $<-19.5$ and $> -19.5$) --- in order to get around
the extinction effect discussed in the previous section --- and compute 
the mean values of $r_{pe}$ and $a_{pe}$, now identified by a subscript
``1''; those values are listed in columns 5 and 6 of Table 1, for each 
luminosity class. The exclusion of galaxies with $hr_{pe}<1$ kpc biases 
high the values of $<hr_{pe}>_1$; it does so however equally for all 
luminosity classes, through which $<hr_{pe}>_1$  remains approximately 
constant. The mean values $<a_{pe}>_1$, which increase with decreasing 
luminosity, illustrate the fact that less luminous populations are on the
average more nearby. Note that the values of $<a_{pe}>_1$ are fairly large 
in comparison with the typical seeing, excluding the likelihood of an 
angular resolution bias.

Next, we inquire on the implications of a constant $r_{pe}$ on the total 
mass distribution. Assume the total mass of the galaxy $M_{200}$ is that 
comprised within a limiting radius $r_{200}$, defined as that 
within which the halo mean density is 200 times the critical density of 
the Universe. Then,
\be
M(r_{pe})=M_{200}{\bar\rho(r_{pe})\over\bar\rho_{200}}
\Bigl({r_{pe}\over r_{200}}\Bigr)^3
\ee 
where $M(r_{pe})$ is the mass within $r_{pe}$, and $\bar\rho(r_{pe})$
and $\bar\rho_{200}$ are respectively the mean density within $r_{pe}$ and
within $r_{200}$. Since $V^2(r) = GM(r)/r$, we can also express the ratio 
between the rotational velocity at $r_{pe}$ and the asymptotic value 
$V_{200}$ as
\be
\Bigl[{V(r_{pe})\over V_{200}}\Bigr]^2 = {\bar\rho(r_{pe})\over\bar\rho_{200}}
\Bigl({r_{pe}\over r_{200}}\Bigr)^2
\ee
For a flat rotation curve, the definition of $r_{pe}$ yields 
$[V(r_{pe})/ V_{200}]^2\simeq 0.4$; then
\be
M(r_{pe}) \simeq 0.4 M_{200}{r_{pe}\over r_{200}}
\ee
It can be shown that $r_{200} \propto M_{200}^{1/3}$; thus, for a constant
value of $r_{pe}$: 
\be
M(r_{pe}) \propto M_{200}^{2/3}
\ee
Since $V_\circ (1+\alpha)$ does not generally coincide with $V_{200}$,
the effect of the luminosity dependence of $\alpha$ needs to be taken
into account: it can be parametrized by writing
$(V(r_{pe})/ V_{200})^2\propto M_{200}^\beta$, where $\beta$ is small
and positive. In general, Equation (6) can then be rewritten as
\be
M(r_{pe}) \propto M_{200}^{2/3+\beta}
\ee

We now inquire on the dependence of the halo mass within $r_{pe}$
on the total mass. We assume a halo density form as proposed by
Navarro, Frenk \& White (1997):
\be
\rho_h(r) = \rho_{crit}{\delta_\circ\over (r/r_s)(1+r/r_s)^2}
\ee
where $\rho_{crit}$ is the cosmological critical density, and the
scaling parameters $\delta_\circ$ and $r_s$ can be expressed in terms
of $r_{200}$ and a concentration index $c\equiv r_{200}/r_s$, with
$\delta_\circ = c^3(200/3)[\ln(1+c)-c/(1+c)]^{-1}$. For small values 
of $r$, the halo mass within radius $r$ is
\be
M_h(r) = {400\pi c^2 \rho_{crit} r_{200}\over \ln(1+c)-c/(1+c)}~r^2
\ee
where $M_{200} \simeq M_h(r_{200})$. For a constant $r_{pe}$, 
\be
M_h(r_{pe}) \propto {c^2\over \ln(1+c)-c/(1+c)} M_{200}^{1/3}
\ee
Numerical simulations indicate that $c$ is mildly dependent on halo mass, 
{\it decreasing} approximately as $(M_{200})^b$ with $b\simeq -0.10$
(Navarro, Frenk \& White 1997). A comparison of Equations (7) and (10)
thus indicates that the mass in the halo cannot account for the inner shape 
of the rotation curve, unless the concentration index {\it increases} with 
halo mass, opposite to what numerical simulations suggest. An alternative 
interpretation is that the contribution of disk and bulge to $M(r_{pe})$ 
increases with galaxy luminosity (and mass), a well--exercised idea in the 
current literature (e.g. Burstein \& Rubin 1985; Broeils 1992; Moriondo, 
Giovanardi \& Hunt 1998; Sellwood 1999 and refs. therein). If we assume that 
$M(r_{pe})=M_h(r_{pe})(1+f_d)$, $\beta\sim 0$ 
and neglect the dependence of $c$ on $M_{200}$,
\be
1+f_d \propto M_{200}^{1/3},
\ee
suggesting that as the galaxy mass increases, the fraction of 
baryonic material in the inner few kpc grows rather quickly. For these 
systems, the baryonic mass fraction may saturate at its cosmological level 
(i.e. these systems retain all their baryons) and their inner rotation curves 
may be completely determined by it.  

The spirals in our sample have small bulges (they are mostly of type Sbc and 
Sc). Within that context, note that in an exponential disk--dominated rotation 
curve a rotational velocity $1-1/e=0.63$ of the maximum is obtained at a radial 
distance of about 0.6 disk scale lengths. For galaxies in our high luminosity 
bin, the mean ratio between $r_{pe}$ and the scale length of the disk light is 
$0.63\pm0.02$, as shown in Table 1, while it rises to twice that value for 
the less luminous galaxies in our sample. 

We conclude that the simplest and most likely explanation of the near constancy 
of $<r_{pe}>$ among luminosity classes is that the baryonic mass fraction within
the inner few kpc of spirals increases with total luminosity, and that within
that region the most luminous spirals are entirely dominated by baryonic matter. 
We integrate photometric profiles to $r=r_{pe}$ in order to obtain
I band luminosities within that radius, assuming a solar absolute magnitude 
of +3.94 (Livingston 2000). Estimating mass from $M(r_{pe})=r_{pe}V(r_{pe})^2/G$, 
for galaxies brighter than $M_I= -22+5\log h$ we obtain a mean baryonic 
mass--to--light ratio $\bar \Upsilon_I=(2.7\pm0.12)h$ ~$M_\odot/L_{\odot,I}$ 
within $r_{pe}$. Individual colors for galaxies in our sample are not
available; however, if we assume that spiral galaxies radiate approximately
like blackbodies with $R-I\sim 0.5$, then a mean bolometric $\bar \Upsilon 
\simeq 1.25 \bar\Upsilon_I = 3.4h$ ~$M_\odot/L_{\odot}$ can be estimated. The
coarse analysis presented here does not take into consideration the
dynamical effect of a central bulge with a low spin parameter and we have
not attempted to separate disk and bulge contributions to the light. It is
interesting to compare our result with that of Bottema (1999) for the
Sb galaxy NGC 7331: from measurements of stellar velocity dispersions
he obtains $\Upsilon_{I,disk}=1.6\pm 0.7$ and $\Upsilon_{I,bulge}=6.8\pm 1.0$.

\section{Conclusions}

We have shown that the \hal/[\ion{N}{2}] rotation curves of spiral galaxies 
exhibit a dependence of the inner slope of their rotation curves on the 
galaxy axial ratio. We have interpreted this result as induced by extinction 
occurring within disks. The effect is luminosity dependent, being stronger 
for the most luminous systems as previously indicated by Giovanelli \etal (1995)
on photometric grounds.

Serendipitously, it was also found that rotation curves of galaxies of
different luminosity classes exhibit approximately the same average
exponential scale length in the inner regions, of about $1.7 h^{-1}$ kpc
(the result has potential in the determination of redshift--independent 
distances: we will report elsewhere on this application). 
This is in agreement with the idea that in the inner parts of spiral
galaxies the baryonic component contributes a rapidly increasing fraction
of the mass, with increasing total luminosity of systems. In fact, the most 
luminous galaxies in our sample appear to be completely baryon--dominated
within the inner few kpc. We use this inference and the combination
of spectroscopic and photometric data to estimate a mean baryonic 
mass--to--light ratio for that region $\bar \Upsilon_I=2.7h$ 
$M_\odot/L_{\odot,I}$ in the I band.

\vskip 0.3in

This work has been supported by NSF grants AST96--17069 and AST99--00695.
Discussions with M. S. Roberts are thankfully acknowledged.

\newpage

\vfill
\end{document}